# Competing Energetics Govern Gas Permeation in Polymer of Intrinsic Microporosity (PIM) Membranes

Jianhao Qian [a], Ruoyu Wang [a] and Menachem Elimelech [a,b*]

[a] Department of Civil and Environmental Engineering, Rice University, Houston, TX, 77005, USA

[b] Rice Center for Membrane Excellence (RiCeME), Rice University, Houston, TX, 77005, USA

[*] Email: menachem.elimelech@rice.edu

## ABSTRACT

Polymer membranes, particularly polymers of intrinsic microporosity (PIMs), hold great promise for gas separation applications. However, the long-dominant solution-diffusion model, which treats the membrane as a nonporous homogeneous medium, does not resolve how gas–solid atomic interactions govern molecular transport in intrinsic micropores, limiting rational bottom-up membrane design. In this work, we employ non-equilibrium molecular dynamics simulations to investigate the permeation of various gases (He, $H_2$, $CH_4$, $N_2$, $O_2$, and $CO_2$) through PIM-1 as a representative PIM membrane across a range of temperatures. By analyzing the scaling of gas permeability with molecular mass, we identify a temperature-induced transition in the dominant transport mechanism. We demonstrate that this transition is governed by the competition between gas-wall interaction potential energy and thermal kinetic energy: weak interactions or elevated temperatures facilitate Knudsen-type ballistic transport, whereas strong interactions and lower temperatures favor adsorption-mediated surface diffusion. Furthermore, molecular trajectory analysis at the membrane interface reveals two distinct entry pathways: direct entry through pore openings and surface-diffusion-assisted entry. The surface-diffusion-assisted pathway greatly promotes the entry of strongly interacting gases into the membrane, contributing to higher overall permeability, albeit this enhancement diminishes with increasing temperature. These findings offer a mechanistic picture of gas permeation in PIM-1 and explain the dependence of gas permeation on both gas type and temperature. More broadly, they highlight the importance of adopting a pore-flow perspective to understand gas transport in microporous polymer membranes.

## I. INTRODUCTION

Membrane-based gas separation has emerged as a more energy-efficient and modular alternative to conventional thermal and chemical separation processes[1], with broad applications including carbon capture[2], hydrogen purification[3], hydrocarbon separation[4], and natural gas sweetening[5]. Despite the emergence of many novel porous membranes[6], polymer membranes remain the dominant material platform in this field, owing to their low cost, scalable fabrication, and tunable chemical structure[1].

Polymer membranes, however, still face several persistent challenges. Trade-offs between permeability and selectivity fundamentally limit separation performance[7]. Physical aging causes membranes to lose porosity and shrink pore size over time, leading to a continuous decline in permeability[8]. Plasticization by condensable gases disrupts chain packing and compromises selectivity[9]. Extensive efforts have been made to address these challenges through various approaches, such as crosslinking[9], polymer blending[10], and large-scale material screening[11]. These studies largely focused on membrane materials, while the molecular mechanisms of gas transport, which is central to membrane gas separation performance, remain less clearly resolved. In particular, a more fundamental question remains overlooked: how gas-membrane interactions at the atomic scale govern gas transport, and how these interactions ultimately determine gas permeability and selectivity. Without this understanding, rational bottom-up design remains elusive.

Gas transport in polymer membranes has long been described by the classic solution-diffusion model[12], which treats the membrane as a nonporous, homogeneous medium and macroscopically decomposes transport into three steps: dissolution into the membrane feed side, diffusion through it, and desorption from the permeate side. This simplification, however, does not explicitly resolve the atomistic mechanisms governing gas molecule transport. Polymer membranes are not truly nonporous and homogeneous. Interchain spacing and chain packing defects form tortuous, pore-like pathways through which molecules migrate[13,14]. Importantly, many emerging polymer membrane designs deliberately increase fractional free volume or introduce more connected transport pathways, further underscoring the central role of pore structure and pore connectivity in gas permeation. Such confined transport of gas molecules through the membrane, along with the resulting molecular sieving effect, has been widely observed in both simulations[15] and experiments[16]. This pore-flow perspective has already proven fruitful in guiding the bottom-up design of zeolite, metal-organic frameworks, and carbon molecular sieve membranes by tailoring pore size and pore wall chemistry[4,17]. Yet for polymer membranes, such a pore flow understanding has received comparatively less attention, especially for distinguishing how different pore-mediated transport mechanisms emerge under different gas–membrane interactions.

Polymers of intrinsic microporosity (PIMs), which achieve permanent porosity through rigid, contorted backbone geometries that prevent efficient chain packing, make a compelling target for a pore-level transport picture[18]. Among them, PIM-1 stands out as a benchmark system, combining exceptional permeability with tractable synthesis and well-characterized pore structure[19]. Recent molecular dynamics studies have provided valuable insights into PIM-1[20,21], yet these works have often focused on macroscopic metrics like solubility and diffusivity within the solution-diffusion model. In such microporous systems, gas transport deviates from continuum diffusion and is increasingly governed by gas-wall interactions[22]. When these interactions are weak relative to thermal kinetic energy, molecules travel in ballistic trajectories between successive wall collisions, a regime known as Knudsen diffusion[23]. When interactions are strong, adsorption is favorable and molecules instead migrate along the pore wall via adsorption-mediated surface diffusion[24]. The dominant gas transport mechanism thus reflects a competition between gas-wall affinity and molecular kinetic energy, which varies with both gas species and temperature[25,26]. While this competition has been discussed at the macroscopic level in rigid inorganic porous materials[27], its atomic-level quantification within the complex, amorphous network of microporous polymers like PIM-1 remains largely unexplored. Therefore, how these mechanisms coexist, compete, and transition in PIM-1 requires detailed investigation.

In this study, we systematically investigate how gas transport mechanisms in PIM-1 change with different gas types and temperature. Non-equilibrium molecular dynamics (NEMD) simulations are employed to analyze the permeability of multiple gas species across different temperatures. We reveal a clear transition in transport mechanisms with temperature through permeability–mass scaling relationship. We then uncover the microscopic origin of this transition, showing that the competition between gas-wall interaction potential energy and thermal kinetic energy governs whether gases undergo adsorption-mediated surface diffusion or Knudsen-type ballistic transport within the pores. This competition not only affects transport within the pores, but also governs how gas molecules enter the membrane, thereby influencing gas uptake into the membrane. Overall, our findings offer a coherent mechanistic picture of gas permeation in PIM-1 and provide molecular-level insights for the design of high-performance microporous polymer membranes.

## II. MOLECULAR DYNAMICS SIMULATION METHODOLOGY

All MD simulations were performed using the LAMMPS package[28]. The OPLS-AA force field[29] was used for PIM-1, He, $CH_4$, and parameters generated by LigParGen[30]. Force field parameters for $H_2$ (ref.[31]), $O_2$ (ref.[31]), $N_2$ (ref.[31]), and $CO_2$ (ref.[32]) were adopted from the literature. A timestep of 1 fs was used throughout all simulations. A cutoff distance of 12 Å was applied to the Lennard-Jones and short-range electrostatic interactions, with long-range electrostatic interactions treated by the Particle-Particle Particle-Mesh (PPPM) method at an accuracy of $1.0 \times 10^{-4}$. Lorentz-Berthelot mixing rules were adopted for Lennard-Jones interactions between atoms with different types. Unless otherwise specified, the canonical (NVT) and isothermal-isobaric (NPT) ensembles were used for temperature and pressure control. The simulation details for each component of this work are described below.

**PIM-1 membrane preparation:** A PIM-1 membrane was constructed from 16 chains, each containing 32 repeat units. The chains were placed in a simulation box of 68 Å × 68 Å in the XY plane with periodic boundary conditions and sufficient length in the Z direction. Two pistons, oriented parallel to the XY plane, were placed at both ends of the box along the Z direction to control pressure by adjusting the applied forces on the pistons. The membrane was then relaxed using a 21-step protocol[33]. A single cycle of 21 steps yields a fresh PIM-1 membrane, while repeating the cycle multiple times produces membranes with higher density and smaller porosity, corresponding to an aged PIM-1 membrane[34]. In this work, only a single 21-step cycle was performed, yielding a fresh PIM-1 membrane with a density of approximately 0.95 g/cm$^3$. After relaxation, the pistons were removed.

**Gas permeability calculation:** A feed side, 20 nm long in the Z direction, and a permeate side were established on either side of the PIM-1 membrane (Fig. 1a). The pressure on the feed side was maintained by controlling the number of gas molecules. When the count dropped, new molecules were inserted to restore the target pressure. Any molecule that crossed the membrane into the permeate side was deleted to maintain vacuum conditions, and the cumulative number of deleted molecules was recorded as the permeate count. To prevent membrane displacement along the pressure gradient, 5% of the atoms within a 1 nm slice of the membrane adjacent to the permeate side were fixed, simulating a support layer. Gas entry into the membrane and transport through most permeation pathways occur without interaction with the fixed atoms, which are sparsely distributed within a thin layer adjacent to the permeate side.

At the start of the simulation, the membrane contained no gas. As the simulation progressed, gas gradually sorbed into the membrane until saturation was reached (Fig. 1b). The cumulative permeate count increased over time and became linear after 10 ns (Fig. 1c). The subsequent 5 ns of simulation data, where the permeate count varied linearly with time, were used to extract the permeability ($P$) by fitting to

$$P = \frac{1}{N_A} \frac{dN}{dt} \frac{L}{A \Delta p} \tag{1}$$

where $N$ is the cumulative permeate count, $N_A$ is the Avogadro number, $t$ is time, $L$ is the membrane thickness, $A$ is the cross-sectional area of the membrane, and $\Delta p$ is the pressure difference across the membrane. In the 5 ns fitting window, the net increase in permeate count was at least 58 molecules and the linear fit $R^2$ exceeded 0.97 for all cases, with most cases above 0.99, confirming sufficient statistics.

In this work, permeability is reported in Barrer, where 1 Barrer = $3.35 \times 10^{-16}$ mol · m / (m$^2$ · s · Pa). A pressure difference of 20 bar was chosen because it produces enough permeation events for meaningful statistical analysis within accessible simulation timescales, and it is also widely used in experimental membrane permeability studies[35]. For He, $H_2$, $N_2$, $O_2$, and $CH_4$ permeating through PIM-1 at near-ambient temperatures, permeability shows little dependence on pressure, while $CO_2$ shows a slight decline with increasing pressure at room temperature[36]. Although $CO_2$ shows a slight pressure dependence near room temperature, this dependence becomes negligible at 500 K, the temperature used for our main analysis (Fig. S1). This confirms that 20 bar is a reasonable driving pressure for our simulation conditions.

**Distribution of distances from pore space to nearest membrane atoms:** A three-dimensional grid with a spacing of 0.5 Å was constructed over the simulation box. For each grid point, the distance to the nearest membrane atom center was calculated. Atomic radii were not considered; therefore, the calculated distance represents the distance from the grid point to the nearest atom center rather than to the atomic surface.

**Gas molecule density distribution within the membrane:** A system consisting of 8 PIM-1 chains (32 repeat units each) and 200 gas molecules was placed in a cubic periodic simulation box. The 21-step protocol was applied to obtain a stable gas-sorbed PIM-1 structure, followed by an NVT simulation at 300 K for 1 ns, with frames saved every 0.01 ns. For each frame, two distance distributions were computed: (i) the nearest distance from each gas molecule atom to membrane atoms, giving the number of gas molecule atoms at distance $d$ from the membrane, $N(d)$, and (ii) the nearest distance from 1000 randomly placed points to membrane atoms, giving the geometrically available volume at distance d, $V(d)$. Dividing $\mathrm{N}(d)$ by $\mathrm{V}(d)$ yields the local gas density $\rho(d)$, which was then normalized so that its integral equals unity, allowing direct comparison across different gas species and temperatures.

**Interaction potential energy of gas adsorption and desorption in the membrane:** The relaxed PIM-1 structure containing gas molecules, obtained from the calculated gas density distribution inside the membrane, was used as the starting configuration. PIM-1 atoms were fixed, and the gas molecules were thermostated to an extremely low temperature (0.1 K), at which kinetic energy is negligible and all gas molecules adsorb onto pore walls. The resulting interaction potential energy between the gas molecules and membrane atoms was then taken as the adsorption energy. To approximate a desorbed state, the $\sigma$ parameter in the Lennard-Jones force field of the gas atoms was artificially increased to gradually inflate the gas molecules, displacing them away from pore walls (Fig. S2). The force field parameters were then restored, leaving gas molecules positioned in the center of pores in the desorbed state, and the corresponding interaction potential energy was taken as the desorption energy. We note that this procedure is a computational construction designed to generate a desorbed configuration, rather than a representation of a physical desorption pathway.

**Gas entry permeance calculation:** As gas molecules migrate from the feed side into the membrane, they first cross a low-density interfacial region before entering the denser interior. Gas molecules were defined as having completed entry when they reached a region where the membrane density equals the average interior density. The region beyond this point was treated as part of the membrane interior. The flux across this interface was used to calculate the so-called entry permeance, following the same approach used for permeability, based on the first 250 permeation events after the system reached steady state (Fig. S3). To accurately capture the back-and-forth reversal events during gas entry, molecular trajectories were recorded with a high time resolution: every 20 fs for He and $H_2$, and every 100 fs for $N_2$, $O_2$, $CH_4$, and $CO_2$. This ensures a sub-angstrom average displacement between consecutive frames. The sensitivity of this frame interval was tested, and the results confirm that this choice is acceptable (Fig. S4).

## III. RESULTS AND DISCUSSION

### A. Temperature-induced transition in transport regimes observed from permeability scaling

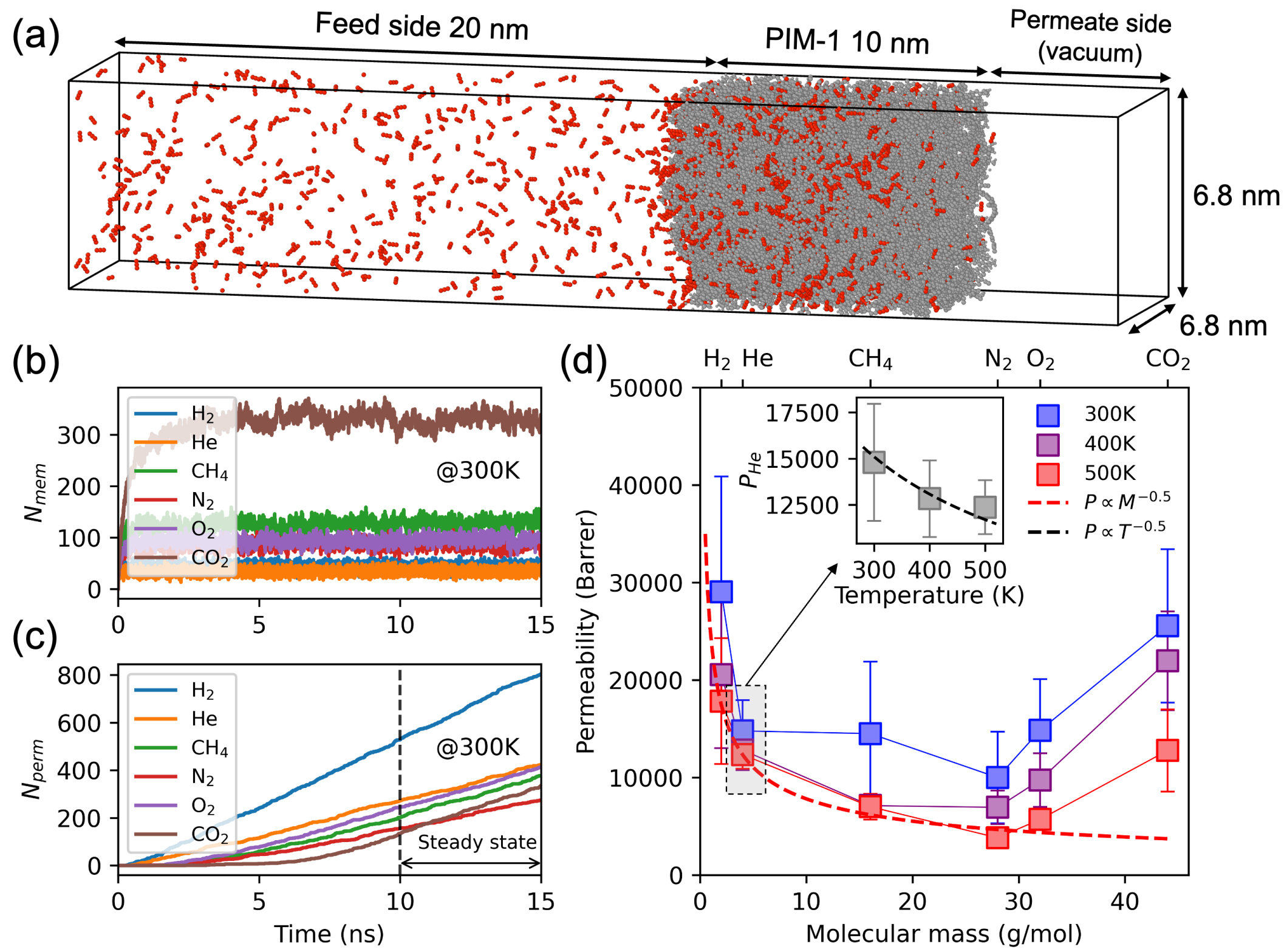

**FIG. 1. Gas permeability through PIM-1 membranes obtained from NEMD simulations.** (a) Snapshot of CO2 permeation through the PIM-1 membrane at steady state with a feed pressure of about 20 bar. Red represents $CO_2$ and gray represents PIM-1. (b) Time evolution of the number of gas molecules within the PIM-1 membrane during the NEMD permeation, showing an initial increase followed by reaching a steady state. (c) Cumulative number of gas molecules exiting at the permeate side as a function of time. (d) Gas permeability of various species ($H_2$, He, $CH_4$, $N_2$, $O_2$, and $CO_2$) through PIM-1 at 20 bar and different temperatures (300 K, 400 K, and 500 K). Error bars represent the standard deviation of three independent PIM-1 simulations. The corresponding data are listed in Table S1.

An effective way to distinguish Knudsen-type transport from adsorption-mediated surface diffusion is to examine how permeability scales with molecular mass. In Knudsen-type transport[23], gas molecules travel back and forth between pore walls; thus, transport is governed by the molecular thermal velocity, which scales as $M^{-0.5}$. This directly leads to a permeability that follows the same $M^{-0.5}$ dependence, meaning lighter molecules permeate faster. In surface diffusion, molecules adsorb onto pore walls and migrate along the surface, where gas-wall interactions become the dominant factor rather than molecular mass. As a result, permeability shows a weaker dependence on $M$. Therefore, measuring the permeability of gases with different molecular masses and checking whether it scales as $M^{-0.5}$ provides a reliable way to assess the contribution of Knudsen-like transport. Because permeability reflects both molecular mobility and gas uptake, permeability–mass scaling alone cannot uniquely determine the transport mechanism. In this study, we therefore use the scaling analysis as an initial diagnostic and combine it with spatial distributions, interaction-energy differences, and molecular trajectories to identify the dominant transport mechanism.

NEMD simulations were performed to obtain permeability data for various gases (see Methods for details). Here, the simulated permeabilities are higher than experimental measurements because our models represent fresh PIM-1, rather than samples that have undergone long-term physical aging and structural compaction. The generated PIM-1 replicas have slightly different porosities, leading to sample-to-sample variation in absolute permeability values and thus relatively large error bars. At 300 K, gas permeability in PIM-1 shows no obvious dependence on molecular mass, suggesting that Knudsen-type transport is not the dominant mechanism for all gases. Furthermore, the lack of dependence on molecular size suggests that molecular sieving is not the primary factor controlling the observed permeability trend. The weak mass and size dependences instead indicate that adsorption-mediated surface diffusion contributes substantially to transport at 300 K.

The temperature dependence of permeability provides an additional signature of the transport mechanism. In a rigid pore, temperature influences permeability primarily through two factors: feed-side gas density and molecular velocity. For an ideal gas, the feed-side density is inversely proportional to temperature ($T^{-1}$); hence, increasing temperature reduces the driving concentration and suppresses permeability. Molecular velocity, on the other hand, scales as the square root of temperature and affects permeability differently depending on the transport mechanism.

In molecular sieving, higher temperatures give molecules enough kinetic energy to pass through narrow pores that are comparable in size to the gas molecules themselves. This activation effect significantly enhances permeability, leading to an Arrhenius-type increase with temperature, as observed in highly crosslinked polyamides[15,37]. In Knudsen-type transport, gas molecules bounce back and forth between pore walls without being trapped at adsorption sites, making permeability directly proportional to molecular velocity. Because velocity scales with $T^{0.5}$ and feed-side gas density depends on $T^{-1}$, the permeability, which is proportional to both[23], is expected to follow a $T^{-0.5}$ dependence. For surface diffusion, elevated temperatures weaken gas adsorption on pore walls, thereby gradually shifting transport toward Knudsen-type transport.

NEMD simulations were performed at elevated temperatures of 400 K and 500 K (Fig. 1d). Among all gases studied, helium permeability exhibits a clear $T^{-0.5}$ temperature dependence, indicating that its transport in PIM-1 is governed by Knudsen-type transport (inset of Fig. 1d). Other gases show a more pronounced decline in permeability with increasing temperature. This further argues against molecular sieving as the dominant mechanism, as this mechanism predicts flux to increase with temperature. Since helium transport follows Knudsen-type transport, we used its permeability as a baseline to predict the Knudsen permeability of other gases using the relation $P \propto M^{-0.5}$. At high temperatures, the simulation results align well with this scaling, indicating a shift toward Knudsen-type transport. This scaling holds consistently across six independent membrane replicas (Fig. S5). One notable exception is $CO_2$, whose permeability drops but remains significantly above the Knudsen prediction, suggesting that surface diffusion continues to contribute even at elevated temperatures.

We compared our simulated gas permeabilities with reported experimental measurements, specifically choosing methanol-treated PIM-1 because this treatment effectively removes residual solvents from the polymer[38]. Our simulated permeabilities are higher than the experimental values (Table S2) due to the lower density of our model and its correspondingly larger pores. The larger pore size increases the permeabilities of all gases by a similar factor, resulting in simulated selectivities that remain close to the experimental values. $CO_2$ is an exception, as a large part of its permeability originates from surface diffusion, which does not scale with pore volume, and its selectivity is therefore lower than measured.

By analyzing the dependence of permeability on molecular mass and temperature, we identify a temperature-induced shift in the dominant gas transport mechanism in PIM-1. However, two fundamental questions remain: why does this shift occur in PIM-1, and why does surface diffusion lead to enhanced permeability?

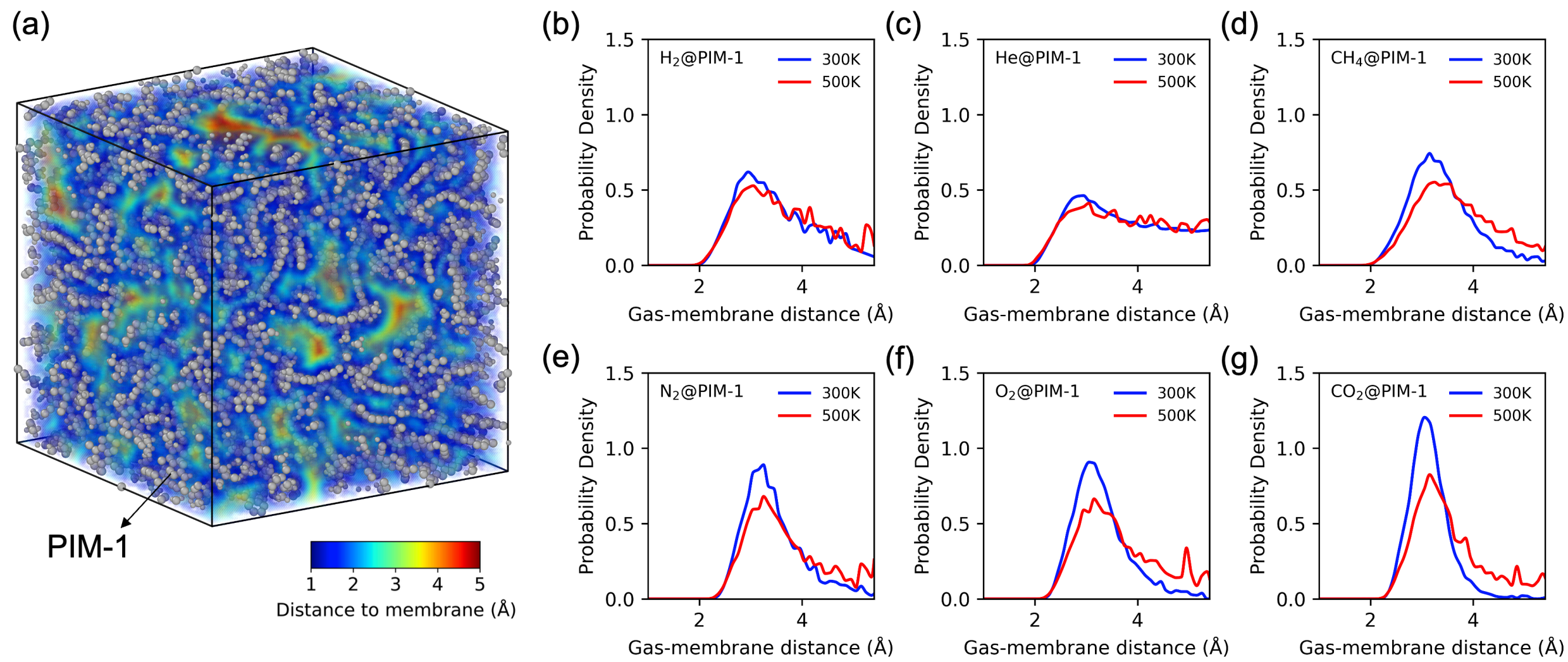

**FIG. 2. Gas distribution within PIM-1 pores.** (a) Spatial distribution of the pore space in the PIM-1 membrane. Membrane atoms are represented by gray spheres. The pore space is colored according to the distance to the nearest membrane atom. (b-g) Density distribution profiles of (b) $H_2$, (c) He, (d) $CH_4$, (e) $N_2$, (f) $O_2$, and (g) $CO_2$ in PIM-1 pores at 300 K and 500 K from EMD simulations.

To further support the transition in transport mechanisms, we analyzed the spatial distribution of gas molecules inside PIM-1 (see Methods for details). In PIM-1, gas molecules within the pores can be located either close to or away from the membrane atoms (Fig. 2a). In Knudsen-type transport, gas molecules are distributed more uniformly across the pore, while in surface diffusion, they tend to accumulate near the membrane atoms. As shown in the density distribution profiles (Fig. 2b–g), strongly interacting gases such as $CO_2$ and $O_2$ exhibit pronounced accumulation near the pore walls at 300 K, with probability density peaks significantly higher than those of weakly interacting gases such as He and $H_2$. This further supports adsorption-mediated surface diffusion for $O_2$ and $CO_2$ in PIM-1, while He and $H_2$ are more inclined toward Knudsen-type transport. At 500 K, the density profiles of all gases become markedly flatter, suggesting a reduced tendency for surface adsorption and a shift toward Knudsen-type transport. An exception is $CO_2$, which still displays a pronounced peak, indicating that surface diffusion remains the dominant transport mechanism. These observations are consistent with the temperature-dependent permeability trends observed (Fig. 1d).

### B. In-pore transport mechanisms governed by kinetic and potential energies

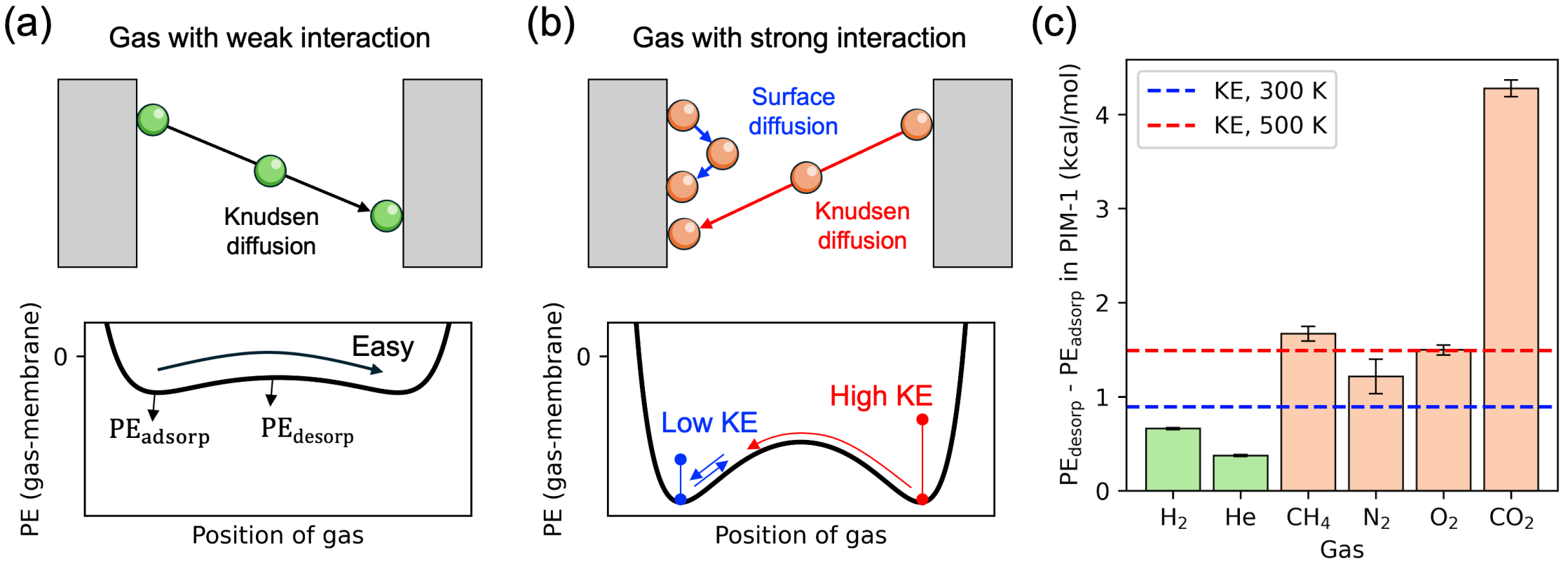

**FIG. 3. Energetic control of in-pore gas transport mechanisms.** (a) Knudsen-type ballistic transport in pores for weakly interacting gases. Gray blocks represent the pore walls, and green spheres represent gas molecules. The lower schematic panel illustrates the interaction potential energy profile across the pore, featuring a shallow energy well that can be readily overcome by thermal motion. Due to the small pore size, gas molecules in the desorbed state still experience residual wall interactions, resulting in a negative potential energy. (b) Transition between adsorption-mediated surface diffusion and Knudsen-type transport for gases strongly interacting with pore walls. Orange spheres denote gas molecules with strong gas-wall interactions. The lower schematic panel illustrates the potential energy profile across the pore, featuring a deeper energy well that confines molecules near the pore wall at low kinetic energy but can be overcome at high kinetic energy. (c) Interaction potential energy differences between the adsorbed and desorbed states for various gases in PIM-1 pores, compared with gas kinetic energies at 300 K and 500 K. $H_2$ and He are classified as weakly interacting gases (green bars), while $CH_4$, $N_2$, $O_2$, and $CO_2$ are categorized as strongly interacting gases (orange bars). Error bars represent the standard deviation of three independent PIM-1 simulations. This comparison is based on average kinetic and interaction energies and therefore provides a qualitative indicator rather than an absolute transition criterion. Because both quantities exhibit statistical distributions, transient adsorption and surface diffusion can still occur for certain molecules even when the average kinetic energy exceeds the potential barrier.

To elucidate the mechanistic origin of gas transport in PIM-1 membranes, including how it is governed by gas-wall interaction and temperature, and how the transition between Knudsen-type transport and surface diffusion occurs, we examined the competition between gas–wall interaction potential energy and thermal kinetic energy. As a gas molecule approaches the pore wall, attractive van der Waals and electrostatic interactions cause the interaction potential energy to decrease gradually, while short-range repulsion increases the potential energy at closer distances, forming a potential energy well that favors adsorption near the pore wall. When a gas molecule is positioned near the pore center, it is in a relatively desorbed state and experiences a less favorable interaction potential energy. The energy difference between adsorbed and desorbed configurations influences the transport mechanism. When gas-wall interactions are weak, this energy difference is small and can be readily overcome by thermal motion of gas molecules, so molecules bounce between pore walls without being trapped at the surface, resulting in Knudsen-type transport (Fig. 3a). Conversely, when gas-wall interactions are strong, the energy difference becomes large (Fig. 3b). At low temperatures, the kinetic energy of gas molecules is insufficient to overcome this energetic preference for the pore wall, confining them to near-wall trajectories and making adsorption-mediated surface diffusion the dominant transport mechanism. As temperature increases, the kinetic energy of gas molecules grows, enabling them to escape the potential well and traverse the pore center, thereby activating Knudsen-type transport.

Such competition between gas-wall interaction potential energy and gas kinetic energy helps explain the mechanistic origin of gas transport in PIM-1. Here, we calculated the adsorption ($\mathrm{PE_{adsorp}}$) and desorption potential energies ($\mathrm{PE_{desorp}}$) between the gas molecules and PIM-1 (see Methods for details), and obtained the potential energy difference ($\mathrm{PE_{desorp}} - \mathrm{PE_{adsorp}}$) shown in Fig. 3c. The thermal translational kinetic energy was calculated as $1.5k_\mathrm{B}T$, where $k_\mathrm{B}$ is the Boltzmann constant and $T$ is the absolute temperature. We did not include rotational and vibrational kinetic energies because they do not move the center of mass or overcome potential energy barriers, making them ineffective for gas transport[25,26]. For $H_2$ and He, the energy differences fall below the kinetic energy at 300 K, indicating that these molecules can easily overcome the adsorption-desorption energy difference at this temperature, resulting in Knudsen-type transport. For $CH_4$, $N_2$, $O_2$, and $CO_2$, the interaction-energy differences are comparatively large, exceeding the kinetic energy at 300 K, which favors near-wall adsorption and promotes adsorption-mediated surface diffusion. As temperature rises to 500 K, the higher kinetic energy enables $N_2$, $O_2$, and $CH_4$ molecules to overcome their interaction-energy differences, shifting their transport from surface diffusion toward Knudsen-type transport. However, for $CO_2$, its exceptionally high interaction-energy difference remains above the kinetic energy even at 500 K; thus, adsorption-mediated surface diffusion continues to dominate $CO_2$ transport at this temperature.

It is worth noting that the comparison above is based on average kinetic and potential energies, which provides a useful but simplified picture. In practice, both kinetic and potential energies follow statistical distributions, meaning that even when the average kinetic energy exceeds the average interaction-energy difference, some gas molecules may still lack sufficient instantaneous kinetic energy to escape local adsorption sites, resulting in transient adsorption and surface diffusion. Therefore, the condition of average kinetic energy exceeding the average interaction-energy difference does not correspond to purely Knudsen-type transport, nor does falling below it imply purely surface diffusion. In general, both mechanisms coexist to different extents[39]. In addition, this potential-energy difference does not account for entropic contributions to desorption. Therefore, its comparison with $3k_\mathrm{B}T/2$ should therefore be interpreted as indicating trends among gases rather than as a quantitative criterion.

### C. Gas entry mechanisms at the membrane interface

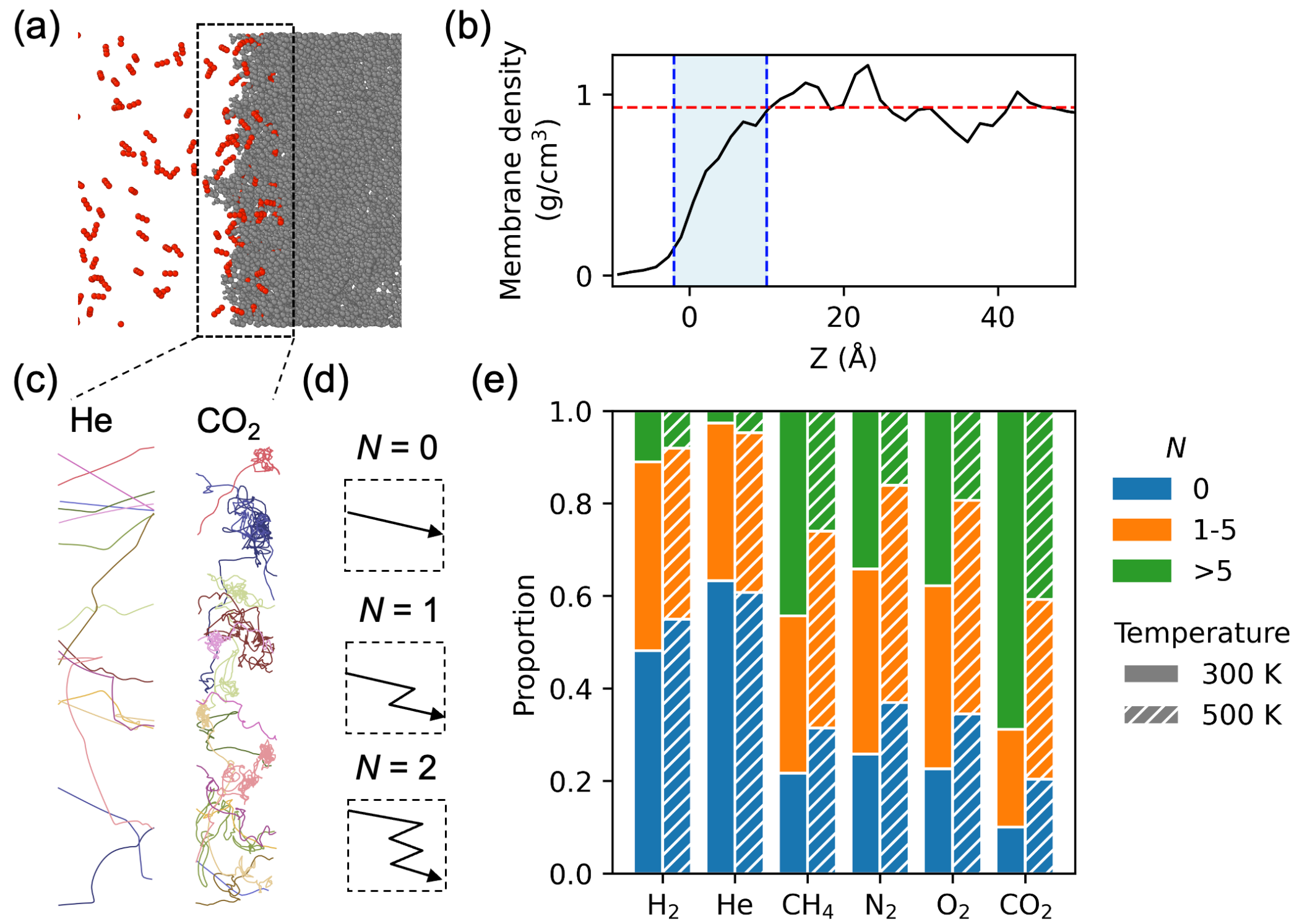

**FIG. 4. Molecular dynamics of gas entry into the PIM-1 membrane.** (a) NEMD snapshot illustrating $CO_2$ entering PIM-1. Red represents $CO_2$ and gray represents PIM-1. The dashed box indicates the entry region. (b) Density profile of the PIM-1 membrane. The red dashed line indicates the average membrane density in the interior region, and the blue dashed lines define the entry region. (c) Representative trajectories of He and $CO_2$ during entry into PIM-1, captured within the entry region defined in (a) and (b) from NEMD simulations (see the Methods section for details). (d) Definition of the number of back-and-forth reversals undergone by gas molecules during membrane entry. (e) Distribution of reversal counts for different gases entering PIM-1 at different temperatures.

From the analysis above, we understand the underlying mechanisms of Knudsen-type transport and adsorption-mediated surface diffusion in PIM-1 membranes, as well as the transitions between them. However, one question remains unclear: Given that strong adsorption traps gas molecules at adsorption sites and slows diffusion within the membrane, why does surface diffusion still lead to an enhanced overall gas permeability (Fig. 1d)? To address this, we examine the flux of gas entering the membrane (Fig. 4a), which complements in-pore transport in determining permeability.

To gain a clearer picture, we first examined how gas molecules enter the membrane under different transport mechanisms. NEMD simulations were performed to capture the process of gas molecules entering the PIM-1 membrane (see the Methods section for details). An entry region was defined spanning from the gas phase to the point where the local membrane density reaches that of the membrane interior (Fig. 4b).

Within this entry region, we tracked the trajectory of each gas molecule that fully traversed the region (Fig. 4c) and counted the number of back-and-forth turns along the transport axis, denoted as $N$ (Fig. 4d). For weakly interacting gases, molecules that collide with the solid part of the interface are simply reflected back to the feed side without being adsorbed. These molecules enter the membrane primarily through pore openings at the interface, with no reversals ($N = 0$). In contrast, strongly interacting molecules can adsorb onto the solid interface and undergo repeated near-surface motion before reaching a pore opening. For practical trajectory classification, entry events with one or more reversals ($N > 0$) are designated as surface-mediated entry, whereas events with no reversal ($N = 0$) are designated as direct entry. This classification serves as an operational descriptor and does not imply that every reversal corresponds to a distinct surface-diffusion event.

The distribution of reversal counts across all tested gases is summarized in Fig. 4e. For weakly interacting gases ($H_2$ and He), the majority of molecules enter with $N = 0$, indicating the dominance of direct, collision-controlled entry. For strongly interacting gases ($CH_4$, $N_2$, $O_2$, and $CO_2$), a significantly larger fraction of molecules exhibits $N > 0$ at 300 K, indicating that entry via surface diffusion is the dominant pathway. As the temperature increases to 500 K, the proportion of $N = 0$ events increases markedly for these gases, indicating that surface diffusion as an entry pathway becomes less pronounced. This can be explained by the increased kinetic energy at elevated temperatures, which facilitates escape from adsorption sites and reduces the likelihood of surface-mediated entry.

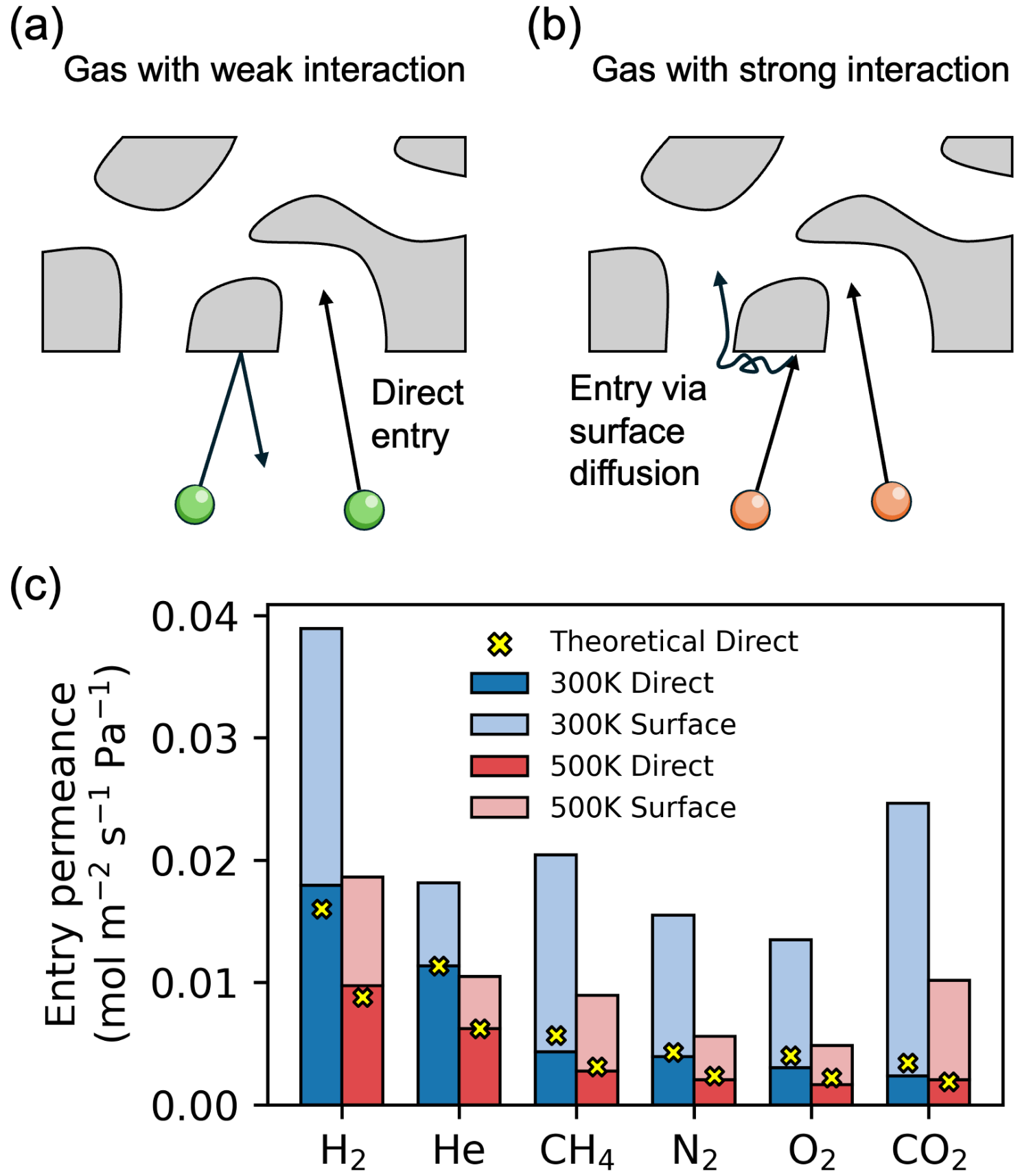

**FIG. 5. Influence of gas-wall interaction strength on gas entry pathways.** (a) Entry pathways for weakly interacting gases into a porous material. Gray blocks represent the porous material, green spheres denote weakly interacting gas molecules, and arrows indicate gas trajectories. Weakly interacting gases are reflected back to the feed side when colliding with the solid surface. They can only enter the membrane by passing directly through interfacial pore openings, as shown by the straight trajectory. (b) Entry pathways for strongly interacting gases into a porous material, where orange spheres represent gas molecules. These gases may adsorb onto the wall when colliding with the solid surface and subsequently enter the membrane via surface diffusion along the boundary toward pore openings. (c) Entry permeance of various gases in PIM-1 at 300 K and 500 K, decomposed into direct entry and surface-diffusion-assisted entry. Yellow crosses represent the direct entry permeance predicted by kinetic theory.

The trajectory analysis reveals that gas molecules can enter PIM-1 through two distinct pathways: direct entry through pore openings and surface-diffusion-assisted entry. These two different transport pathways have also been observed in graphene nanopores[40]. Strongly interacting gases benefit from an additional surface-mediated entry route that is not available to weakly interacting gases (Fig. 5a and 5b). To quantify the relative contributions of these two pathways, we first calculated the total entry permeance of each gas (see the Methods section for details). By analyzing the entry trajectories, we then determined the fractions of molecules entering via $N = 0$ and $N > 0$, corresponding to direct entry and surface-diffusion-assisted entry, respectively. Multiplying these fractions by the total entry permeance yields the individual contributions of each pathway to the overall entry permeance.

Direct entry is governed by gas kinetic theory. The permeance is proportional to molecular velocity and therefore inversely proportional to the square root of molecular mass. Using helium as a reference, the direct permeance of other gases can be estimated as:

$$P_{\mathrm{direct},i}^{\mathrm{th}} = P_{\mathrm{direct,He}} \times \sqrt{\frac{M_{\mathrm{He}}}{M_i}} \quad (2)$$

We compared the direct entry permeance obtained from NEMD simulations with predictions from kinetic theory and found good agreement (Fig. 5c), supporting the validity of the entry pathway classification. For weakly interacting gases, direct entry accounts for the majority of membrane entry. For strongly interacting gases, however, this fraction is negligibly small, particularly at low temperatures.

At 300 K, gases such as $CH_4$, $N_2$, $O_2$, and $CO_2$ exhibit entry permeance values comparable to that of He, despite being much heavier. The entry permeance of these strongly interacting gases is predominantly contributed by the surface diffusion pathway. Due to energy competition, as temperature increases to 500 K, the kinetic energy of gas molecules increases, rendering the adsorption energy wells insufficient to trap and adsorb them. This weakens surface diffusion and reduces entry via this pathway, leading to a marked decrease in the entry permeance of gases that rely predominantly on surface diffusion for membrane entry.

The analysis of surface-diffusion-assisted entry explains two key observations in PIM-1 permeability (Fig. 1d): (i) strongly interacting gases can exhibit higher permeability than weakly interacting gases because they access an additional surface-mediated entry pathway that boosts entry permeance, and (ii) their permeability drops more sharply at elevated temperatures because increased kinetic energy at elevated temperatures suppresses adsorption and reduces the contribution of surface-mediated entry. These two entry pathways also provide a microscopic reading of the solubility term in the solution-diffusion model. A gas that reaches pore openings through adsorption and surface diffusion has an additional mechanism for entering the membrane, increasing its uptake and therefore its apparent solubility. Because adsorption becomes less favorable at higher temperatures, this pathway contributes less to membrane uptake, explaining the decrease in solubility with temperature.

## IV. CONCLUSION

In this work, by adopting a molecular-scale pore-flow perspective, we show that, in fresh PIM-1, gas permeation is influenced by the physical competition between gas-wall interaction potential energy and thermal kinetic energy. Within the pores, this energetic balance drives a temperature-induced transport transition: weak interactions and elevated temperatures facilitate Knudsen-type ballistic transport, whereas strong gas-wall affinities at lower temperatures confine molecules to adsorption-mediated surface diffusion. Crucially, this same energy competition governs the interfacial dynamics, revealing two distinct membrane-entry pathways: direct pore entry and surface-diffusion-assisted entry. The surface-diffusion-assisted pathway is associated with the elevated entry permeance of strongly interacting gases, consistent with their high overall permeability. As thermal kinetic energy increases, this surface-assisted entry is systematically suppressed.

By elucidating the competing energetics and dual entry pathways, this work provides a molecular-level, mechanism-based interpretation of gas transport in microporous polymers that complements, while going beyond, the phenomenological macroscopic description provided by solution-diffusion model. For example, while the high $CO_2$ permeability of PIM-1 is well understood to come from high solubility within the $P = S \times D$ framework, our findings show how the favorable gas-polymer interactions underlying this solubility are manifested at the membrane interface through surface-diffusion-assisted entry, and how this process is affected by temperature.

Although the present results are demonstrated for fresh PIM-1, the identified competition between gas–wall interaction potential energy and thermal kinetic energy is expected to be relevant to other microporous polymer membranes. More generally, these findings suggest that gas transport in polymer membranes can be interpreted within a pore-mediated framework in which adsorption and kinetic effects jointly determine the dominant transport mechanism.

## SUPPLEMENTARY MATERIAL

The supplementary material includes additional details on the simulation methodology for generating desorbed gas configurations (Fig. S1), the pressure dependence of $CO_2$ permeability at 500 K (Fig. S2), simulated gas permeabilities through PIM-1 at different temperatures (Table S1), and a comparison of experimental and simulated gas permeabilities and selectivities for PIM-1 (Table S2).

## ACKNOWLEDGMENTS

This research was supported by computational resources from Rice University's Center for Research Computing, as well as the Texas Advanced Computing Center (TACC) on Stampede3 accessed through the Advanced Cyberinfrastructure Coordination Ecosystem: Services and Support (ACCESS) program (Grant No. EVE250006).

Supplementary Materials for

# Competing Energetics Govern Gas Permeation in Polymer of Intrinsic Microporosity (PIM) Membranes

Jianhao Qian [a], Ruoyu Wang [a] and Menachem Elimelech [a,b*]

[a] Department of Civil and Environmental Engineering, Rice University, Houston, TX, 77005, USA

[b] Rice Center for Membrane Excellence (RiCeME), Rice University, Houston, TX, 77005, USA

[*] Email: menachem.elimelech@rice.edu

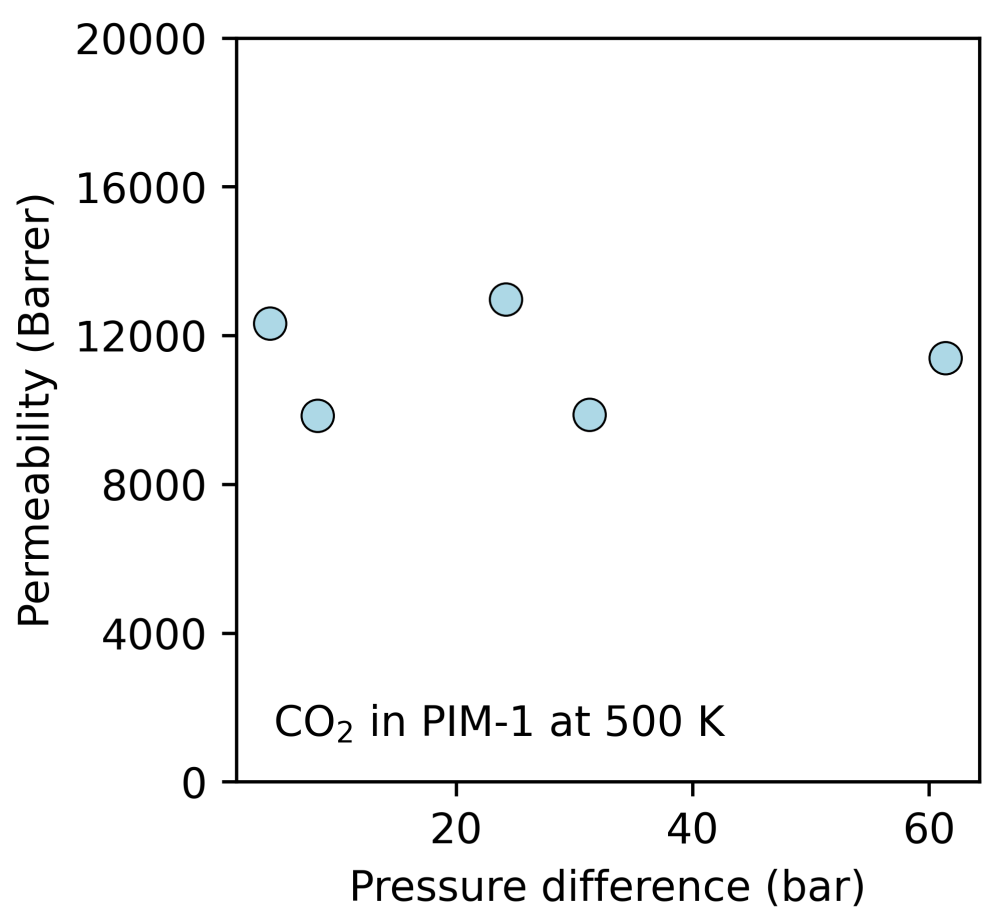

**Figure S1.** Permeability of $CO_2$ in a PIM-1 sample at 500 K as a function of pressure difference.

[Alt text description] Scatter plot of CO2 permeability in PIM-1 at 500 K against pressure difference. The data points stay roughly constant across the tested pressure range, indicating that permeability is nearly independent of driving pressure at this temperature.

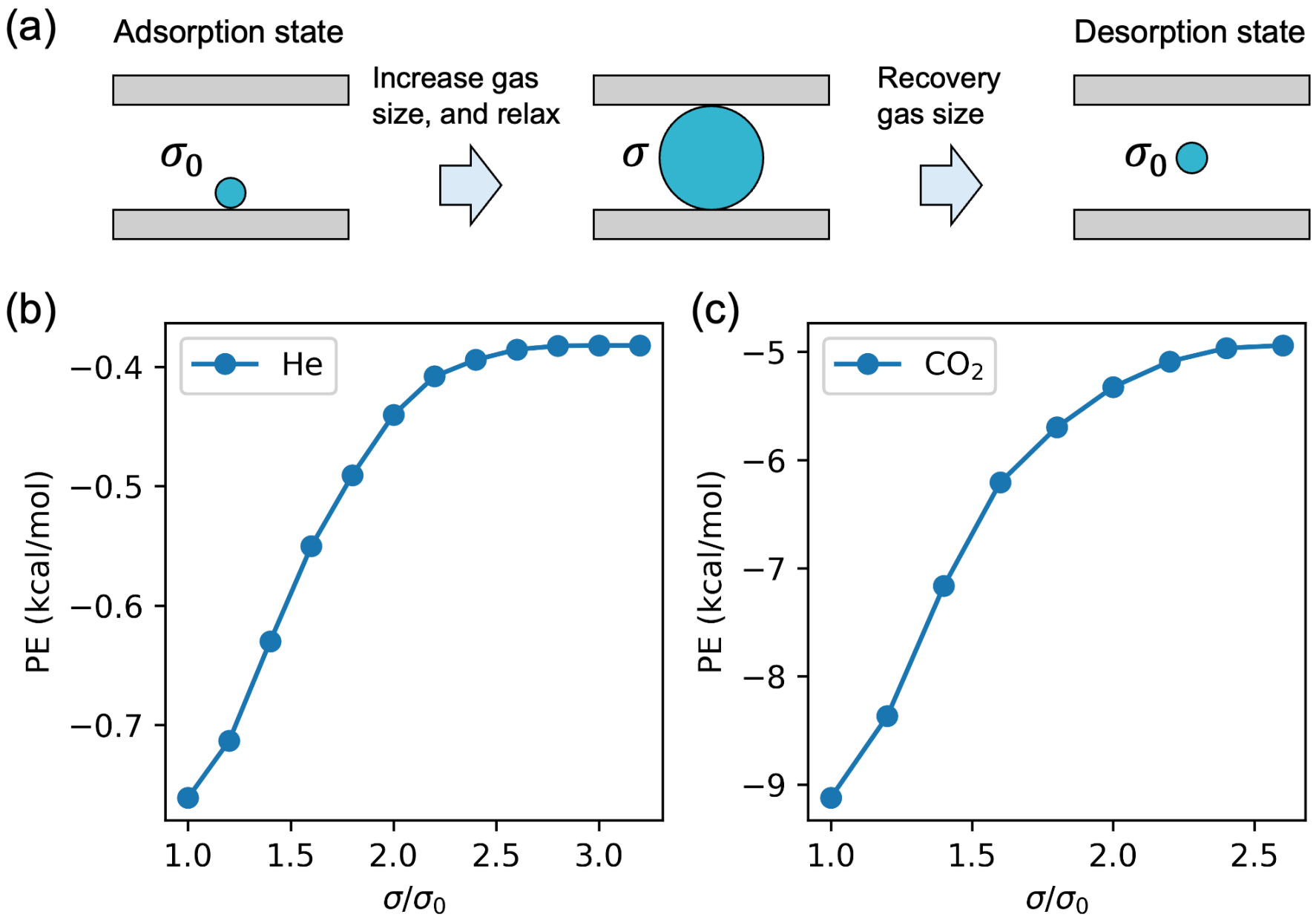

**Figure S2.** (a) Schematic of the method used to obtain the desorption state of a gas molecule. The blue circle represents the gas molecule and the gray blocks represent the pore walls. The Lennard-Jones parameter $\sigma$, which represents the gas size, is first increased artificially to push the molecule out of its adsorption site, then restored to its original value $\sigma_0$, giving the desorption configuration. (b, c) Potential energy (PE) between the gas and PIM-1 as a function of $\sigma/\sigma_0$ for He and CO2, respectively. The PE converges as $\sigma/\sigma_0$ increases.

[Alt text description] Schematic and two line plots illustrating how a desorbed gas configuration is generated. The schematic shows a gas molecule pushed away from the pore wall by artificially enlarging its size, then restored to give the desorption state. Two line plots show the gas-PIM-1 potential energy for helium and CO2 rising and converging as the size ratio increases.

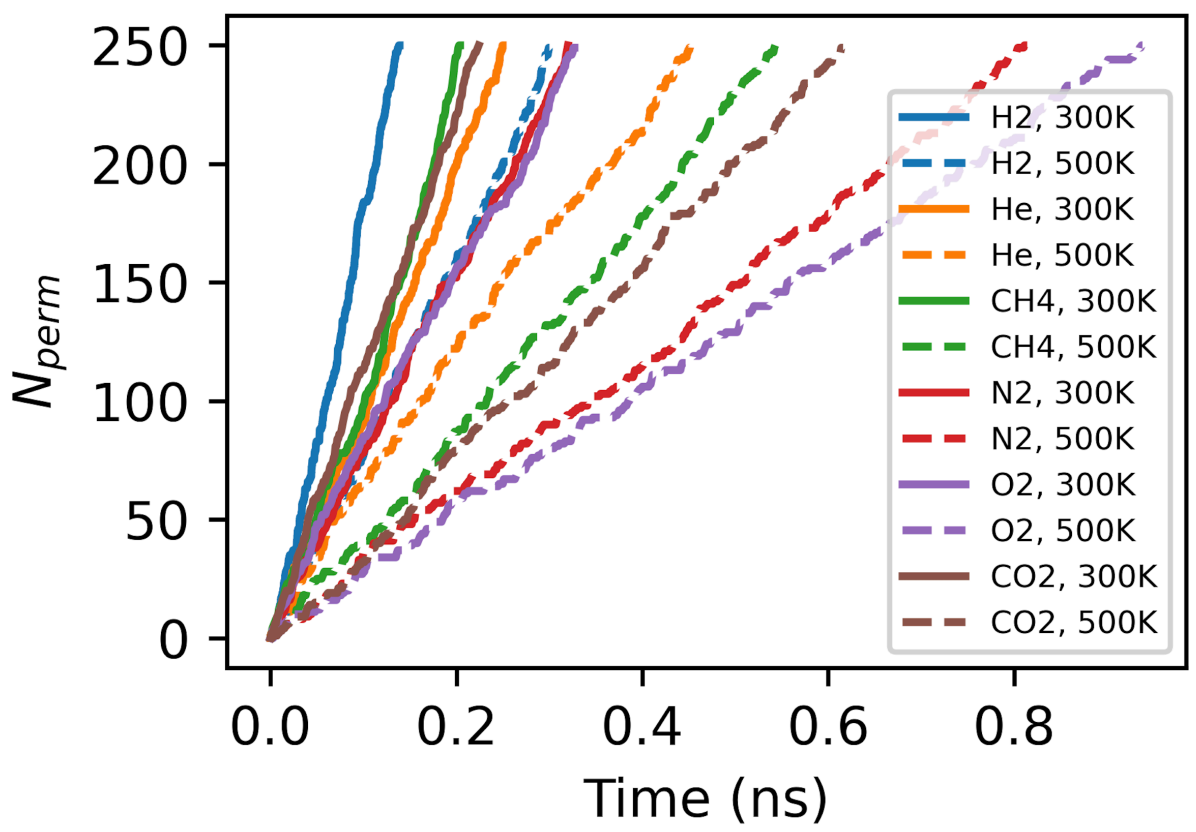

**Figure S3.** Cumulative number of permeation events through the entry region as a function of time for different gases at 300 K and 500 K.

[Alt text description] Line plot showing cumulative permeation events over time for six gases at 300 K and 500 K. All curves rise linearly from zero to 250 events, with 300 K curves (solid) rising faster than 500 K curves (dashed).

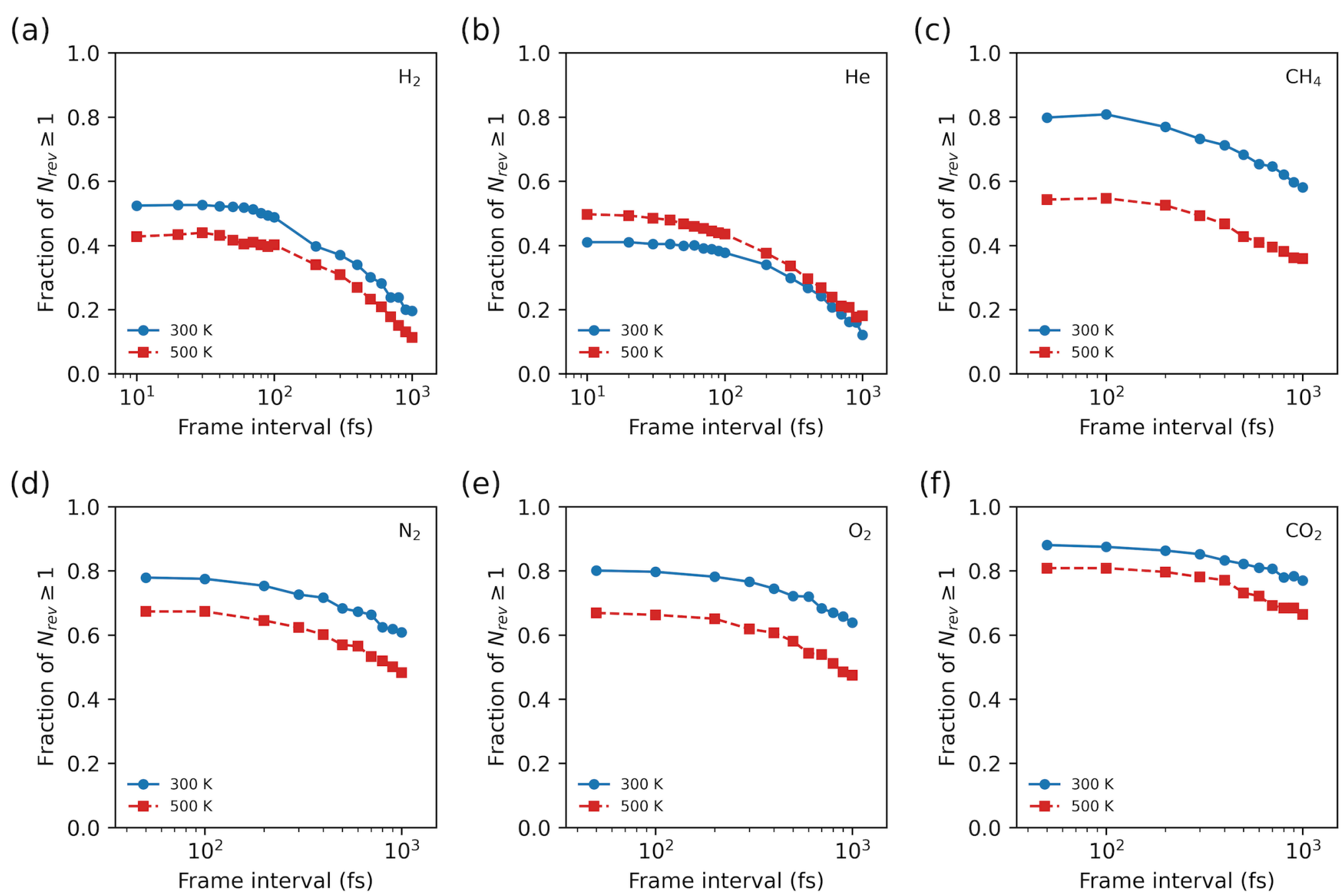

**Figure S4.** Fraction of gas molecules classified as surface-diffusion-assisted entry ($N_{rev} \geq 1$) as a function of the trajectory frame interval, for (a) $H_2$, (b) He, (c) $CH_4$, (d) $N_2$, (e) $O_2$, and (f) $CO_2$ at 300 K and 500 K. The fraction is nearly constant near the frame intervals used in this work, 20 fs for $H_2$ and He and 100 fs for the other gases, showing that the classification is not sensitive to the trajectory resolution.

[Alt text description] Six panels showing the fraction of surface-diffusion-assisted entry versus frame interval, one per gas, each with 300 K and 500 K curves. All curves stay roughly flat at short intervals and decline at longer intervals.

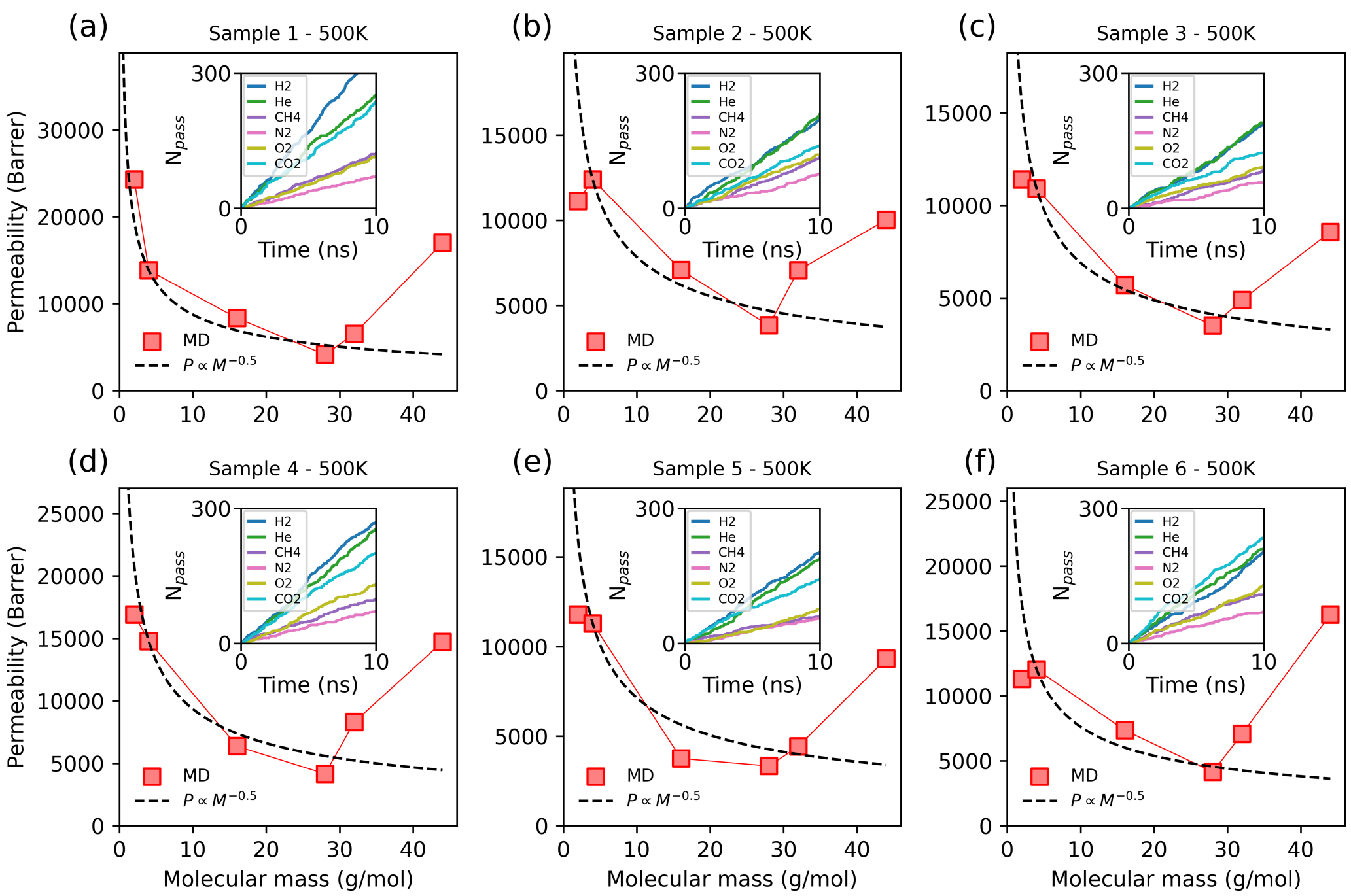

**Figure S5.** Permeability as a function of molecular mass for gases in each PIM-1 sample at 500 K. The inset shows the number of permeated molecules used to calculate permeability as a function of simulation time, counted after the system had reached equilibration. Because of differences in porosity, the permeability values differ across samples. However, the permeability follows $M^{-0.5}$ scaling well for gases except $CO_2$.

[Alt text description] Six panels showing permeability versus molecular mass for six PIM-1 samples at 500 K, with insets of cumulative permeated molecules over time. Most points follow the $M^{-0.5}$ dashed curve, with $CO_2$ (rightmost point) falling above it.

**Table S1.** Simulated gas permeabilities through PIM-1 at different temperatures. Values are reported in Barrer as mean and standard deviation (n = 3).

| Gas | Permeability at 300 K | | Permeability at 400 K | | Permeability at 500 K | |
|---|---|---|---|---|---|---|
| | Ave. | Std. | Ave. | Std. | Ave. | Std. |
| $H_2$ | 29066 | 11804 | 20558 | 7534 | 17847 | 6447 |
| He | 14789 | 3166 | 12833 | 2071 | 12371 | 1463 |
| $CH_4$ | 14513 | 7374 | 7126 | 512 | 7014 | 1326 |
| $N_2$ | 10007 | 4684 | 6953 | 1738 | 3829 | 313 |
| $O_2$ | 14840 | 5264 | 9745 | 2759 | 5707 | 814 |
| $CO_2$ | 25568 | 7865 | 21979 | 5082 | 12782 | 4217 |

[Alt text description] Table listing simulated gas permeabilities (in Barrer) through PIM-1 at 300 K, 400 K, and 500 K for six gases: $H_2$, He, $CH_4$, $N_2$, $O_2$, and $CO_2$. Each entry gives the average and standard deviation from three simulation runs.

**Table S2.** Comparison of experimental and simulated gas permeability and selectivity for PIM-1.

| Gas | Experiments[1] (Standard PIM-1, 303 K) | | Experiments[1] (MeOH treated PIM-1, 303 K) | | Simulations (This work, 300 K) | |
|---|---|---|---|---|---|---|
| | Permeability (Barrer) | Selectivity to He | Permeability (Barrer) | Selectivity to He | Permeability (Barrer) | Selectivity to He |
| He | 740 | 1.00 | 1320 | 1.00 | 14789 | 1.00 |
| $H_2$ | 1600 | 2.16 | 3300 | 2.50 | 29066 | 1.97 |
| $O_2$ | 530 | 0.72 | 1530 | 1.16 | 14839 | 1.00 |
| $N_2$ | 155 | 0.21 | 610 | 0.46 | 10007 | 0.68 |
| $CH_4$ | 240 | 0.32 | 1160 | 0.88 | 14513 | 0.98 |
| $CO_2$ | 3700 | 5.00 | 11200 | 8.48 | 25568 | 1.73 |

[Alt text description] Table comparing experimental and simulated gas permeabilities (in Barrer) and selectivity relative to He for PIM-1. Experimental data are shown for standard and methanol-treated PIM-1 at 303 K. Simulation results are from this work at 300 K. Six gases are included: He, $H_2$, $O_2$, $N_2$, $CH_4$, and $CO_2$.